\documentclass[column,showpacs,preprintnumbers,amsmath,amssymb]{revtex4}
\preprint{}
\usepackage{graphicx}
\usepackage{dcolumn}
\usepackage{bm}
\usepackage{mathrsfs}
\begin{document}
\title{Gravitational time dilation induced decoherence during spontaneous emission}
\author{Dong  Xie}
\email{xiedong@mail.ustc.edu.cn}
\affiliation{Faculty of Science, Guilin University of Aerospace Technology, Guilin, Guangxi, P.R. China.}

\author{Chunling Xu}
\affiliation{Faculty of Science, Guilin University of Aerospace Technology, Guilin, Guangxi, P.R. China.}

\author{An Min Wang}
\email{anmwang@ustc.edu.cn}
\affiliation{Department of Modern Physics , University of Science and Technology of China, Hefei, Anhui, China.}

\begin{abstract}
We investigate decoherence of quantum superpositions induced by gravitational time dilation and spontaneous emission between two atomic levels. It has been shown that gravitational time dilation can be an universal decoherence source.
Here, we consider decoherence induced by gravitational time dilation only in the situation of spontaneous emission. Then, we obtain that the coherence of particle's position state depends on reference frame due to the time dilation changing the distinguishability of emission photon from two positions of particle. Changing the direction of light field can also result in the difference about the coherence of quantum superpositions. For observing the decoherence effect mainly due to gravitational time dilation,  time-delayed feedback can be utilized to increase the decoherence of particle's superpositions.
\end{abstract}

\maketitle

\section{Introduction}
Quantum phenomenon has been observed by numerous experiments on microscopic scales. However, on macroscopic scales, it is difficult to find quantum effects, such as quantum superpositions. A lot of physicists have been looking up the root of quantum-to-classical transition for decades. The reason can be  divided into two categories: coarsened measurement and decoherence\cite{lab1,lab2,lab3,lab4,lab5,lab6,lab7,lab8}.
Commonly viewpoint is that decoherence  plays a prominent role in quantum-to-classical transition. There are two routes to explain decoherence: one route is that system interacts with external environments, the other is taken in wave function collapse\cite{lab9,lab10,lab11}, which need not external environments.
The latter one is often inspired by general relativity and makes a fundamental modification on quantum theory. Recently, Igor at al.\cite{lab12} demonstrated  the existence of decoherence induced by gravitational time dilation without any modification of quantum mechanics. This work motivates further study on decoherence due to time dilation.

Spontaneous emission between two atomic levels inevitably occurs. We research decoherence due to time dilation during spontaneous emission. Without spontaneous emission, decoherence will not occur in our model  only by time dilation. As we all know, spontaneous emission can induce decoherence. We find that gravitational time dilation can reduce or increase the decoherence due to spontaneous emission in different reference frames (different zero potential energy point). It is attributed to the fact that in different reference frame, the distinguishability of emission photon from different positions is different. The direction of emission light also influences the coherence of quantum superpositions in fixed direction of gravitational field.
In order to make the decoherence due to time dilation stronger than due to spontaneous emission, time-delayed feedback control\cite{lab121,lab1211} is used.

The rest of paper is arranged as follows. In section II, we present the model about the decoherence of quantum superpositions due to time dilation during spontaneous emission. Coherence of particle's position in different reference frame is explored in section III. In section IV, we discuss the influence of different directions of emission light. In section V, a
time-delayed feedback scheme is utilized to increase decoherence induced by gravitational time dilation.  We deliver a conclusion and outlook in section VI.
\section{Model}
Firstly, let us simply review the gravitational time dilation which causes clocks to run slower near a massive object. Given a particle of rest mass $m$ with an arbitrary internal Hamiltonian $H_0$, which interacts with the gravitational potential $\Phi(x)$. The total Hamiltonian $H$ is described by\cite{lab12,lab122}
\begin{eqnarray}
H=H_{ext}+H_0[1+\Phi(x)/c^2-p^2/(2m^2c^2)],
\end{eqnarray}
where $H_{ext}$ is external Hamiltonian. For a free particle, $H_{ext}=mc^2+p^2/2m+m\Phi(x)$. In Eq.(1), the last term, $-H_0p^2/(2m^2c^2)$, is simply the velocity-dependent special relativistic time dilation. The coupling with position, $H_0\Phi(x)/c^2$, represents the gravitational time dilation. When we consider slowly moving particles, $p\approx0$, the gravitational time dilation will be the main source of time dilation. It will not be canceled by the velocity-dependent special relativistic time dilation.

We consider that an atom with two levels is in superposition of two vertically distinct positions $x_1$ and $x_2$.
The atom is coupled to a single unidirectional light field, as depicted in Fig. 1.
\begin{figure}[h]
\includegraphics[scale=0.30]{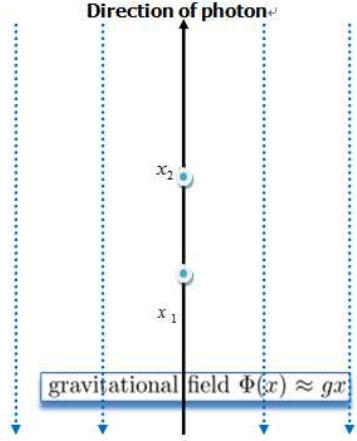}
 \caption{\label{fig.2} Decoherence of an atom is induced by gravitational time dilation under the situation of spontaneous emission. The dotted line represents a homogeneous gravitational field $\Phi(x)\approx g x$, where $g =9.81m/s^2 $ is the gravitational acceleration on Earth. Initial atom is in the superposition state: $1/\sqrt{2}(|x_1\rangle+|x_2\rangle)$. The direction of emitting photon is along the solid line, x-direction, which is contrary with the direction of gravitational field.  }
 \end{figure}

The whole system interacts with a homogeneous gravitational field $\Phi(x)\approx g x$ which generates the gravitational time dilation. The total system-field Hamiltonian is described by ($\hbar=1$)
\begin{eqnarray}
H=[mc^2+m g x_1+w_1(1+g x_1/c^2)|1\rangle\langle1|+w_2(1+g x_1/c^2)|2\rangle\langle2|]|x_1\rangle\langle x_1|\nonumber
\\+\sqrt{\kappa_1/2\pi}(1+g x_1/c^2)|x_1\rangle\langle x_1|\int dw [a b^\dagger(w)\exp(-i w x_1/c)+H.c]\nonumber\\
+[mc^2+m g x_2+ w_1(1+g x_2/c^2)|1\rangle\langle1|+ w_2(1+g x_2/c^2)|2\rangle\langle2|]|x_2\rangle\langle x_2|\nonumber\\
+\sqrt{\kappa_2/2\pi}(1+g x_2/c^2)|x_2\rangle\langle x_2|\int dw [a b^\dagger(w)\exp(-i w x_2/c)+H.c]+\int dw w b^\dagger(w)b(w),
\end{eqnarray}
where $w_1$ and $w_2$ ($w_1>w_2$) are eigenvalues for the atomic level 1 and 2, respectively, and operator $a=|2\rangle\langle1|$.
$\kappa_1$ and $\kappa_2$ denote the coupling constants in position $x_1$ and $x_2$, respectively. Without extra control, the two coupling constants should be same: $\kappa_1=\kappa_2=\kappa$. The last term in Eq.(2) represents the free field Hamiltonian, and the filed modes, $b(w)$, satisfy $[b(w),b^\dagger(w')]=\delta(w-w')$.
Using Pauli operator, $\sigma_z=|1\rangle\langle1|-|2\rangle\langle2|$, to simplify the Eq.(2), we can obtain the new form of system-field Hamiltonian
\begin{eqnarray}
H=[E_1+w_0/2(1+g x_1/c^2)\sigma_z]|x_1\rangle\langle x_1|+\sqrt{\kappa/2\pi}(1+g x_1/c^2)|x_1\rangle\langle x_1|\int dw [a b^\dagger(w)\exp(-i w x_1/c)+H.c]\nonumber\\
+[E_2+w_0/2(1+g x_2/c^2)\sigma_z]|x_2\rangle\langle x_2|+\sqrt{\kappa/2\pi}(1+g x_2/c^2)|x_2\rangle\langle x_2|\int dw [a b^\dagger(w)\exp(-i w x_2/c)+H.c]\nonumber\\+\int dw w b^\dagger(w)b(w),
\end{eqnarray}
where $E_i=mc^2+\frac{(w_1+w_2)}{2}(1+g x_i/c^2)$ for $i=1,2$ and $w_0=w_1-w_2$.

We consider the initial field in the vacuum state and the atom in the state $|1\rangle\frac{|x_1\rangle+|x_2\rangle}{\sqrt{2}}$.
Then, the atom will spontaneously emit photon. According  to there being only a single excitation  conservation between system and field\cite{lab13}, the system state in any time $t$ can be solved analytically, see Appendix.

\section{Coherence of particle's position}
The quantum coherence of particle's position state can be quantified by the interferometric visibility $V(t)$, as shown in Eq.(27) in Appendix. When the time $t$ satisfy $\lambda_1\kappa t\gg1$ and $\lambda_2\kappa t\gg1$, the amplitude of  excitation state $C_1\approx0$ and $C_2\approx0$. Then, we arrive at
\begin{eqnarray}
V=\frac{2\kappa\lambda_1\lambda_2}{\sqrt{[\kappa(\lambda_1^2+\lambda_2^2)]^2+[w_0(\lambda_2-\lambda_1)]^2}}
\exp[-\lambda_1^2\kappa\tau],
\end{eqnarray}
where $\lambda_i=1+g x_i/c^2$ for $i=1,2$.
From the above equation, we can see that the decoherence comes from the spontaneous emission (when $\lambda_1=\lambda_2=1$) and the gravitational time dilation. Spontaneous emission can generate the decoherence due to the fact that photon is emitted from different positions, which leads to having a phase difference $w\tau$, where $w$ denotes the frequency of photon.

And we achieve that coherence depends on the reference frame. Different zero potential energy point (different value of $\lambda_1$ ) will give different coherence strength. The counterintuitive result occurs because in different frame the phase difference will become different so that the distinguishability of emitting photon from two positions is different. Reducing the zero potential point (increasing the value of $\lambda_1$), the phase difference will increase because of time dilation. For a fixed position difference $\Delta=g(x_2-x_1)/c^2$, the quantum coherence can be rewritten
\begin{eqnarray}
V(\lambda_1,\Delta)=\frac{2\kappa\lambda_1(\lambda_1+\Delta)}{\sqrt{[\kappa(\lambda_1^2+(\lambda_1+\Delta)^2)]^2+(w_0\Delta)^2}}
\exp[-\lambda_1^2\kappa\tau].
\end{eqnarray}
There is an optimal value of $\lambda_1$, which can give the maximal quantum coherence, as shown in Fig. 2.
\begin{figure}[h]
\includegraphics[scale=1]{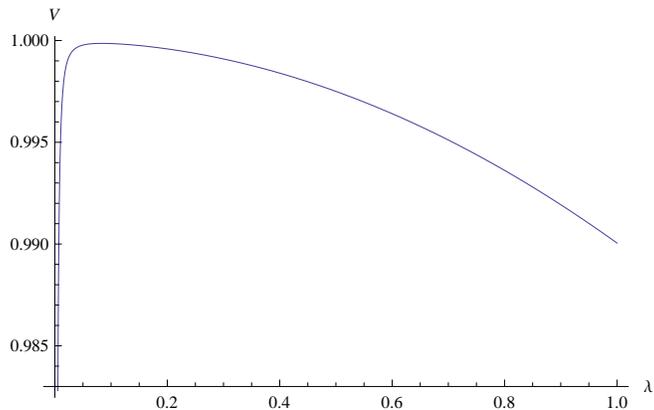}
 \caption{\label{fig.2}Diagram of quantum coherence $V$ changing with $\lambda_1$. The quantum coherence depends on reference frame. The parameters are given by: $w_0/\kappa=10^6$, $\Delta=10^{-6}$, $\kappa \tau=10^{-2}$.}
 \end{figure}
In an optimal reference frame, one can obseve the maximal coherence: $V$ is close to 1.

 In order to observe the decoherence induced by gravitational time dilation, it need to satisfy that the decoherence effect from time dilation is stronger than only from spontaneous emission ($\lambda_1=\lambda_2=1$):
\begin{eqnarray}
\frac{2\kappa\lambda_1(\lambda_1+\Delta)}{\sqrt{[\kappa(\lambda_1^2+(\lambda_1+\Delta)^2)]^2+(w_0\Delta)^2}}\exp[-\lambda_1^2\kappa\tau]
\ll\exp[-\kappa\tau].
_{}\end{eqnarray}
Noting that the value of $\lambda_2-\lambda_1$ is generally small in experiment, the condition $\exp[(\lambda_1^2-1)\kappa\tau]\gg1$ is necessary for observing decoherence mainly induced by gravitational time dilation.

When one changes the direction of emitting photon, the quantum coherence will change accordingly,
\begin{eqnarray}
V'=\frac{2\kappa\lambda_1\lambda_2}{\sqrt{[\kappa(\lambda_1^2+\lambda_2^2)]^2+[w_0(\lambda_2-\lambda_1)]^2}}
\exp[-\lambda_2^2\kappa\tau].
\end{eqnarray}
It is due to that the phase difference changes with the direction of emitting photon, becoming $-w\tau$.  Different directions of emitting photon in the fixed gravitational field will generate different quantum coherence $V$.

Then, we consider general three-dimensional space: the emitting photon can be along any direction, as shown in Fig. 3.
\begin{figure}[h]
\includegraphics[scale=0.40]{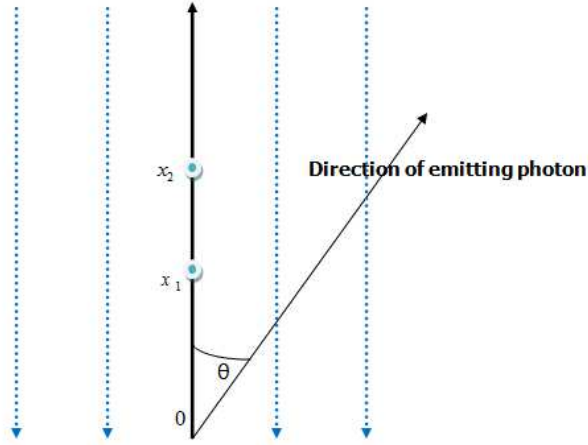}
 \caption{\label{fig.2}Diagram shows that the photon can spontaneously emit in any direction. Here $\theta$ denotes the angle between direction of photon and x-direction, which can change from 0 to $\pi$.}
 \end{figure}
 We obtain the quantum coherence of particle's position state as following:
\begin{eqnarray}
&V_3=\frac{3\kappa\lambda_1\lambda_2}{\sqrt{[\kappa(\lambda_1^2+\lambda_2^2)]^2+[w_0(\lambda_2-\lambda_1)]^2}}|
\int_0^{\pi/2} d\theta\sin\theta\cos^2\theta
\exp[(iw_0\lambda_1-\lambda_1^2\kappa)\tau\cos\theta]+\exp[-(iw_0\lambda_2+\lambda_2^2\kappa)\tau\cos\theta]|,\\
&=\frac{3\kappa\lambda_1\lambda_2}{\sqrt{[\kappa(\lambda_1^2+\lambda_2^2)]^2+[w_0(\lambda_2-\lambda_1)]^2}}|[-2 + \exp(k_1) (2 - 2 k_1 + k_1^2)]/k_1^3+[-2 + \exp(k_2) (2 - 2 k_2 + k_2^2)]/k_2^3|,\\
&\textmd{in which},\nonumber\\
&k_j=[(-1)^jiw_0\lambda_j-\lambda_j^2\kappa]\tau, \ \textmd{for} \ j=1,2,
\end{eqnarray}
where the coupling strength between atom and light field changes with the direction of emitting photon, becoming $\sqrt{\kappa/2}\cos\theta $\cite{lab14}.
For $w_0\lambda_j\tau\ll1$ and $\lambda_j^2\kappa\tau\ll1$, $V_3\approx\frac{2\kappa\lambda_1\lambda_2}{\sqrt{[\kappa(\lambda_1^2+\lambda_2^2)]^2+[w_0(\lambda_2-\lambda_1)]^2}}(1-\lambda_1^2\kappa\tau-\lambda_2^2\kappa\tau)<V'<V.$ It means that the quantum coherence in general three-dimensional space is smaller than in one-dimensional space of fixed direction.
For $w_0\lambda_j\tau\gg1$ and $\lambda_j^2\kappa\tau\gg1$, $V_3\approx\frac{2\kappa\lambda_1\lambda_2}{\sqrt{[\kappa(\lambda_1^2+\lambda_2^2)]^2+[w_0(\lambda_2-\lambda_1)]^2}}|\cos3\varphi|(3/[(w_0^2\lambda_1^2+(\lambda_1^2\kappa\tau)^2]^{3/2}+3/[(w_0^2\lambda_1^2+(\lambda_1^2\kappa\tau)^2]^{3/2})\geq V,$ with $\cos\varphi=\lambda_1^2\kappa\tau/\sqrt{w_0^2\lambda_1^2+(\lambda_1^2\kappa\tau)^2}$. It means that in new condition the quantum coherence in general three-dimensional space is larger than in one-dimensional space of fixed direction.
The root of generating $V_3\neq V$ is the phase difference changing from $w\tau$ to $w\tau\cos\theta$.
\section{time delay feedback}
When one chooses the center of two positions as the zero potential point, the interferometric visibility reads
\begin{eqnarray}
V_c=\frac{2\kappa(1-\Delta/2)(1+\Delta/2)}{\sqrt{[\kappa((1-\Delta/2)^2+(1+\Delta/2)^2)]^2+(w_0\Delta)^2}}
\exp[-(1-\Delta/2)^2\kappa\tau].
\end{eqnarray}
In order to observe the decoherence from gravitational time dilation, not from the spontaneous emission, it is necessary to satisfy  condition $V_c\ll\exp[-\kappa\tau]$. However, the value of $\Delta$ is very small in experiment. So, the condition is hard to meet. We can utilize the time delay feedback \cite{lab15,lab16} to increase the decoherence from gravitational time dilation.

\begin{figure}[h]
\includegraphics[scale=0.30]{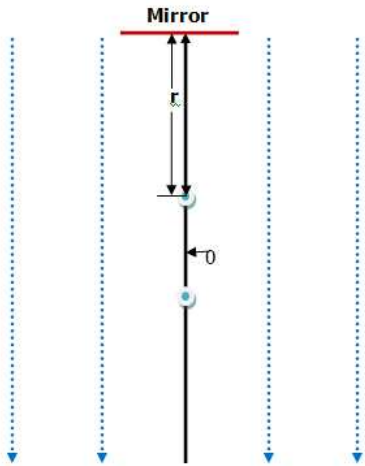}
 \caption{\label{fig.2}Diagram of time delay feedback. The center of two positions is chosen as zero potential point. So, in the new reference frame,  the position $x_1$ ($x_2$) is transformed to be $\Delta c^2/2g$ ($-\Delta c^2/2g$). Here,  we just consider that the photon emits along the fixed x-direction, which can be easily generalized to the case of  three-dimensional space. At $r+\Delta c^2/2g$, a mirror is put to reflect the light field, leading to that a time-delay light field is fed back to system-field interaction.  }
 \end{figure}
As shown in Fig. 4, the light field is reflected by a mirror. The whole system-field Hamiltonian can be described by
\begin{eqnarray}
&H=\sqrt{\kappa/2\pi}(1+g x_1/c^2)|x_1\rangle\langle x_1|\int dw \{a b^\dagger(w)2\exp[-i w(r+\Delta c^2/2g)]\cos(2w r)+H.c\}\nonumber\\
&+\sqrt{\kappa/2\pi}(1+g x_2/c^2)|x_2\rangle\langle x_2|\int dw\{a b^\dagger(w)2\exp[-i w(r+\Delta c^2/2g)]\cos[ w (2r+2\Delta c^2/g)/c]+H.c\}\nonumber\\
&+\int dw w b^\dagger(w)b(w)+[E_1+w_0/2(1+g x_1/c^2)\sigma_z]|x_1\rangle\langle x_1|+[E_2+w_0/2(1+g x_2/c^2)\sigma_z]|x_2\rangle\langle x_2|,
\end{eqnarray}

Using the way in Appendix, we can obtain the quantum coherence at time $t\gg1$. With the feedback, the spontaneous emission is suppressed due to superposition effect.  The total system-field wave function can also be described by Eq.(13) in Appendix.
When the conditions $w_0(1+\Delta/2)2r/c=n\pi$ and $w_0(1-\Delta/2)(2r+2\Delta c^2/g)/c\neq m\pi$ hold, for $t\gg1$ the amplitudes $|C_2|^2=\exp[-n\pi\kappa(1+\Delta/2)/w_0]$ and $|C_1|\simeq0$, where $n,m=1,2,3\cdot\cdot\cdot$. When $w_0\gg\kappa$ and $n=1$, $|C_2|^2\simeq1$. So, we achieve that the quantum coherence $V_c\simeq0$. Without gravitational time dilation, the quantum coherence is much larger than 0. So, utilizing time-delayed feedback scheme can satisfy that the decoherence induced by the gravitational time dilation is far less than by spontaneous emission.
\section{conclusion and outlook}
We explore the decoherence of an atom's positions induced by the gravitational time dilation only in the situation of spontaneous emission. As the phase difference of photon emitted from two positions are different in different reference frames, the quantum coherence of superposition state of positions depends on the reference frame. So one can choose proper reference frame to observe the decoherence from the gravitational time dilation. It is worth mentioning that the direction of emitting photon will influence the quantum coherence. So comparing the case of fixed emitting direction with the case of any direction, there are some differences about quantum coherence. When one chooses the center of two positions as the zero potential point, the decoherence induced by the gravitational time dilation is difficult to be far larger than by spontaneous emission. The time delay feedback can be used to increase the decoherence from the time dilation with proper conditions.

In this article, we only discuss the decoherence of an atom with two energy levels induced by gravitational time dilation. It is interesting to research the decoherence of many particles with many energy levels induced by time dilation with spontaneous emission. In this case we believe that it will increase the decoherence effect from the gravitational time dilation. And considering extra drive is the further research direction. In this situation, due to the fact that a single excitation  conservation between system and field do not hold, the question will become complex and rich.

\section{Acknowledgement}
This work was supported by the National Natural Science Foundation of China under Grant  No. 11375168.

\newpage
                      \ \           $ \mathbf{ APPENDIX}$\\
In the single photon limit, the total system-field wave function is described by
\begin{eqnarray}
|\Psi(t)\rangle=C_1|x_1\rangle|1\rangle|0\rangle+\int dw C_{1w}b^\dagger(w)|x_1\rangle|2\rangle|0\rangle+C_2|x_2\rangle|1\rangle|0\rangle+\int dw C_{2w}b^\dagger(w)|x_2\rangle|2\rangle|0\rangle.
\end{eqnarray}
The variables $C_1$, $C_{1w}$, $C_2$ and $C_{2w}$ denote the corresponding amplitudes of the four states at time $t$.
Applying the Schr$\ddot{o}$dinger equation in the rotating frame, we arrive at the following set of partial differential equations:
\begin{eqnarray}
i\partial_tC_1=[E_1+w_0/2(1+g x_1/c^2)]C_1+(1+g x_1/c^2)\sqrt{\kappa/2\pi}\int dw \exp[i(wx_1/c-w)t]C_{1k},\\
i\partial_tC_{1w}=[E_1-w_0/2(1+g x_1/c^2)]C_{1k}+(1+g x_1/c^2)\sqrt{\kappa/2\pi}\int dw \exp[i(-wx_1/c+w)t]C_{1},\\
i\partial_tC_2=[E_2+w_0/2(1+g x_2/c^2)]C_1+(1+g x_2/c^2)\sqrt{\kappa/2\pi}\int dw \exp[i(wx_2/c-w
)t]C_{2k},\\
i\partial_tC_{2w}=[E_2-w_0/2(1+g x_2/c^2)]C_{2k}+(1+g x_2/c^2)\sqrt{\kappa/2\pi}\int dw \exp[i(-wx_2/c+w)t]C_{2}.
\end{eqnarray}
Substituting $C'_1=\exp[-i(E_1+w_0/2(1+g x_1/c^2))t]C_1, C'_{1k}=\exp[-i(E_1-w_0/2(1+g x_1/c^2))t]C_{1k},$
$C'_2=\exp[-i(E_2+w_0/2(1+g x_2/c^2))t]C_2, C'_{2k}=\exp[-i(E_2-w_0/2(1+g x_2/c^2))t]C_{2k}$ into above equations, we arrive at the following simplified equations:
\begin{eqnarray}
i\partial_tC'_1=(1+g x_1/c^2)\sqrt{\kappa/2\pi}\int dw \exp[i(wx_1/c+(w_0(1+g x_1/c^2)-w)t]C'_{1w},\\
i\partial_tC'_{1w}=(1+g x_1/c^2)\sqrt{\kappa/2\pi}\int dw \exp[i(-wx_1/c+(w-w_0(1+g x_1/c^2))t]C'_{1},\\
i\partial_tC'_2=(1+g x_2/c^2)\sqrt{\kappa/2\pi}\int dw \exp[i(wx_2/c+(w_0(1+g x_2/c^2)-w)t]C'_{2w},\\
i\partial_tC'_{2w}=(1+g x_2/c^2)\sqrt{\kappa/2\pi}\int dw \exp[i(-wx_2/c+(w-w_0(1+g x_2/c^2))t]C'_{2}.
\end{eqnarray}
Eq.(15) and Eq.(17) are integrated formally and inserted into Eq.(14) and Eq.(16),respectively. Utilizing the integral
\begin{eqnarray}
\int dw\exp[i((w_0(1+g x_1/c^2)-w)t]=2\pi\delta(t),
\end{eqnarray}
we can analytically solve the set of partial differential equations.

At time $t$, using the initial values $C_1(0)=1/\sqrt{2}$ and $C_2(0)=1/\sqrt{2}$, we obtain
\begin{eqnarray}
C'_1(t)=1/\sqrt{2}\exp[-1/2\lambda_1^2\kappa t],\\
C'_{1w}=\frac{1-\exp[-1/2\lambda_1^2\kappa t-i(\lambda_1w_0-w)t]}{\lambda_1^2\kappa+i(\lambda_1w_0-w)}\sqrt{\kappa/2\pi}\lambda_1\exp[iwx_1/c],\\
C'_2(t)=1/\sqrt{2}\exp[-1/2\lambda_2^2\kappa t],\\
C'_{2w}=\frac{1-\exp[-1/2\lambda_2^2\kappa t-i(\lambda_2w_0-w)t]}{\lambda_2^2\kappa+i(\lambda_2w_0-w)}\sqrt{\kappa/2\pi}\lambda_2\exp[iwx_2/c],
\end{eqnarray}
where $\lambda_i=1+g x_i/c^2$ for $i=1,2$.
The quantum coherence of position state can be quantified by the interferometric visibility
\begin{eqnarray}
V(t)&=&2|C_1^*C_2+\int dwC^*_{1w}\int dw'C_{2w'}|\nonumber\\
&=&2|\exp[i(x_2-x_1)w_1g/c^2t]{C'_1}^*C'_2+\exp[i(x_2-x_1)w_2g/c^2t]\int dwC'^*_{1w}\int dw'C'_{2w'}|,
\end{eqnarray}
where the term $\int dwC'^*_{1w}\int dw'C'_{2w'}$ can be integrated by residue theorem. We can arrive at
\begin{eqnarray}
&\int dwC'^*_{1w}\int dw'C'_{2w'}=\frac{\kappa\lambda_1\lambda_2}{\kappa(\lambda_1^2+\lambda_2^2)+iw_0(\lambda_2-\lambda_1)}
\{\exp[-1/2\lambda_1^2\kappa\tau+ iw_0\lambda_1\tau]-\exp[-1/2\lambda_1^2\kappa t+i\lambda_1w_0t+i\xi(\tau-t)]-\nonumber\\
&\exp[-1/2\lambda_2^2\kappa t-i\lambda_2w_0t+i(w_0\lambda_1+i\lambda_1^2\kappa)(\tau+t)]+\exp[-1/2\lambda_1^2\kappa (t+\tau)-1/2\lambda_1^2\kappa t+i(\lambda_2-\lambda_1)w_0t+i\lambda_1w_0\tau]\},\\
&\textmd{in which},\nonumber\\
&\tau=(x_2-x_1)/c\geq0,\\
&\textmd{for}\ t<\tau,   \ \xi=w_0\lambda_1+i\lambda_1^2\kappa,\ \ \textmd{for}\ t\geq\tau,\  \xi=w_0\lambda_2-i\lambda_2^2\kappa.
\end{eqnarray}

 \end{document}